\documentclass[prl, 10pt, letterpaper, twocolumn, aps]{revtex4-1}

\usepackage{graphicx,amsmath,amssymb}
\usepackage{color}
\usepackage{ulem}
\usepackage{epstopdf}
\usepackage{lmodern}

\def\Rb{$^{87}$Rb}

\def\ket#1{\left|#1\right\rangle}

\def\ketF#1{\left|#1\right\rangle}

\newcommand{\be}{\begin{equation}}
\newcommand{\ee}{\end{equation}}

\newcommand{\beq}{\begin{eqnarray}}
\newcommand{\eeq}{\end{eqnarray}}

\newcommand{\ops}{\hat{a}_{\rm +1}}      
\newcommand{\opi}{\hat{a}_{\rm -1}}       
\newcommand{\opp}{\hat{a}_{0}}            
\newcommand{\Hrf}{\hat{H}_{\rm rf}}      	              
\newcommand{\Ups}{\hat{U}_{\rm ps}}   	              
\newcommand{\Ubs}{\hat{U}_{\rm bs}}   	              

\newcommand{\gm}{g}
\newcommand{\eem}{\hat{e}}
\newcommand{\gbm}{\asm}

\newcommand{\asm}{h}                                           
\newcommand{\avef}{\bar{f}}                                           

\begin{document}

\hbadness = 10000

\title{0.75 atoms improve the clock signal of 10,000 atoms}

\author{I. Kruse$^1$, K.~Lange$^1$, J. Peise$^1$, B. L\"u{}cke$^1$, L.~Pezz\`e$^2$, J.~Arlt$^3$, W. Ertmer$^1$, C. Lisdat$^4$, L. Santos$^5$, A.~Smerzi$^2$, C. Klempt$^{1}$}

\affiliation{
$^1$Institut f\"ur Quantenoptik, Leibniz Universit\"at Hannover, Welfengarten~1, D-30167~Hannover, Germany \\
$^2$ QSTAR, INO-CNR and LENS, Largo Enrico Fermi 2, I-50125, Firenze, Italy\\
$^3$ Institut for Fysik og Astronomi, Aarhus Universitet, Ny Munkegade 120, DK-8000 \AA{}rhus C, Denmark\\
$^4$ Physikalisch-Technische Bundesanstalt, Bundesallee 100, D-38116 Braunschweig, Germany\\
$^5$ Institut f\"ur Theoretische Physik, Leibniz Universit\"at Hannover, Appelstra\ss{}e~2, D-30167~Hannover, Germany
}

\begin{abstract}
Since the pioneering work of Ramsey, atom interferometers are employed for precision metrology, in particular to measure time and to realize the second.
In a classical interferometer, an ensemble of atoms is prepared in one of the two input states, whereas the second one is left empty. 
In this case, the vacuum noise restricts the precision of the interferometer to the standard quantum limit (SQL).
Here, we propose and experimentally demonstrate a novel clock configuration that surpasses the SQL by squeezing the vacuum in the empty input state. 
We create a squeezed vacuum state containing an average of 0.75 atoms to improve the clock sensitivity of $10,000$ atoms by $2.05^{+.34}_{-.37}\,$dB.
The SQL poses a significant limitation for today's microwave fountain clocks, which serve as the main time reference.
We evaluate the major technical limitations and challenges for devising a next generation of fountain clocks based on atomic squeezed vacuum.
\end{abstract}

\maketitle

Precision measurements allow to probe the boundaries of our understanding of physics.
Prominent recent examples include the discovery of gravitational waves with optical interferometers~\cite{Abbott2016} and the improving bounds on the drift of fundamental constants with atomic interferometers~\cite{Huntemann2014,Godun2014}. 
The current generation of atomic and optical interferometers is however fundamentally limited by vacuum noise, the so-called standard quantum limit (SQL). 
Squeezing the vacuum entering one port of an optical gravitational wave detector was proposed~\cite{Caves1981} in the 80s to surpass the SQL when measuring the {\it length difference} between two optical paths. 
Can squeezed vacuum be useful to improve the measurement of {\it time}? 
Up to now, the concept of vacuum squeezing has not been transferred to atomic clocks or atom interferometry in general.

In this Letter, we design and implement an atom interferometer in clock configuration which exploits atomic squeezed vacuum. 
The clock is operated by combining $N=10^4$ atoms in one input state with a quadrature-squeezed vacuum with an average of 0.75 atoms in the second input state.
The squeezed-vacuum is generated by spin-changing collisions in a Bose-Einstein condensate of neutral $^{87}$Rb atoms -- in direct analogy to optical parametric down-conversion~\cite{Klempt2010,Scherer2010,Luecke2014}.
In contrast to existing methods~\cite{Appel2009,Wasilewski2010,Schleier-smith2010,Bohnet2014,Behbood2014,Hosten2016,Gross2010,Riedel2010,Strobel2014,Luecke2011}
to increase the sensitivity of atomic clocks beyond the SQL in large ensembles, 
our concept disentangles the challenge of increasing the number of atoms (in the main input state) from the creation of squeezing (in the vacuum state).
In particular, the vacuum state remains weakly populated during its preparation, making it immune to losses. 
These central advantages are also exploited in squeezed-vacuum optical interferometers for the detection of gravitational waves, as demonstrated in GEO600~\cite{Schnabel2011} and LIGO~\cite{Aasi2013}, where coherent states of $>10\,$W are combined with a low-power squeezed vacuum state to achieve sub-SQL measurement uncertainty.

\begin{figure*}[ht!]
	\centering
	\includegraphics[width=\textwidth]{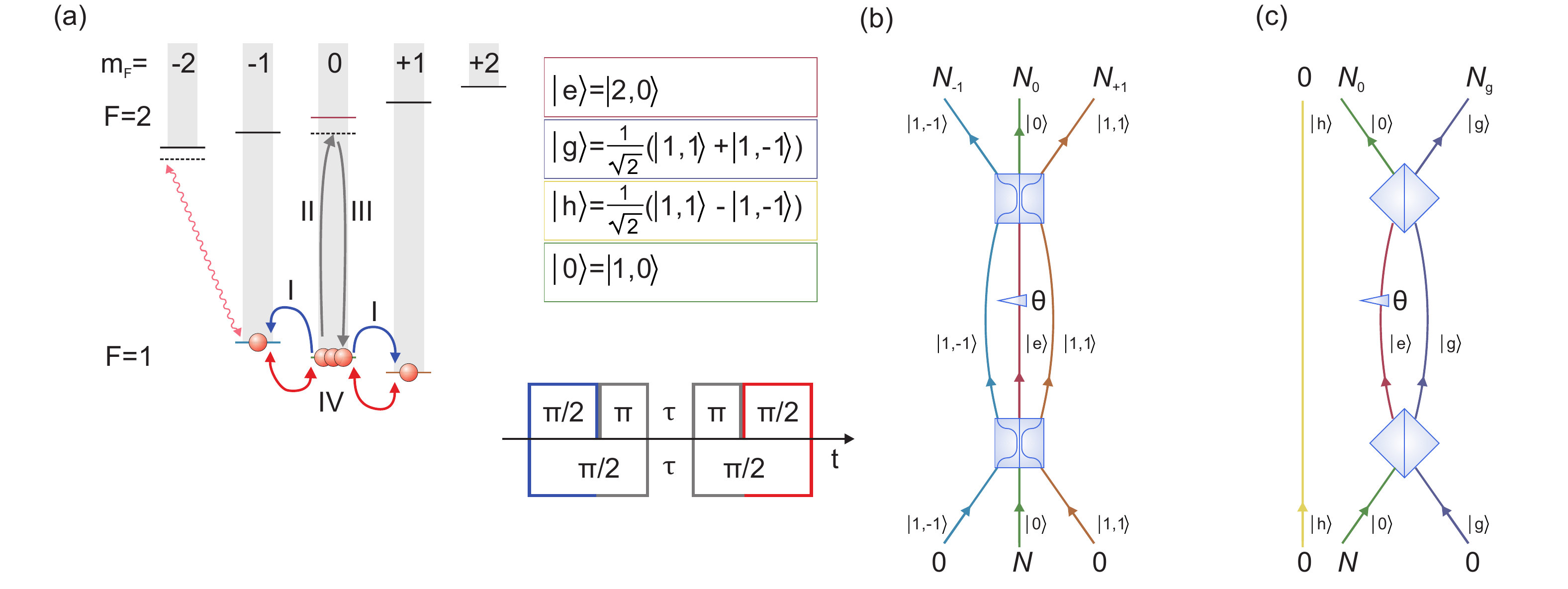}
		\caption{
		\label{fig1}		
		The three-mode interferometer.
		(a) Hyperfine ground states of \Rb.
		(I) A Bose-Einstein condensate in the state $\ket{0}=\ket{F=1,m_F=0}$ is coupled to the states $\ket{1,\pm 1}$ by a resonant radio-frequency pulse.
		The second-order Zeeman shift is compensated by a detuned microwave dressing to the state $\ket{2,-2}$.
		(II) A detuned microwave pulse couples the states $\ket{0}=\ket{1,0}$ and $\ket{e}=\ket{2,0}$.
		These two pulses form an effective $\pi/2$ pulse between $\ket{1,\pm 1}$ and $\ket{e}$.
		(III) After a Ramsey evolution time $\tau_\mathrm{R}$, a second detuned microwave pulse is applied.
		(IV) A final radio-frequency pulse coupling $\ket{0}$ and $\ket{1,\pm 1}$ closes the interferometer.
		(b) The interferometer corresponds to two three-mode beam splitters and a phase shift in between.
		The total number of atoms $N$ enters the central state $\ket{0}$. A measurement of all three components $N_{-1,0,1}$ after the interferometer allows for an estimation of the phase shift $\theta$.
		(c) The interferometer can be simplified by introducing the symmetric state $\ket g$ and the antisymmetric state $\ket{\asm}$.
		The three-mode beam splitters only couple to the symmetric state, thus yielding an effective two-mode interferometer with an unchanged antisymmetric state.}
\end{figure*}

\begin{figure}[ht!]
	\centering
	\includegraphics[width=\columnwidth]{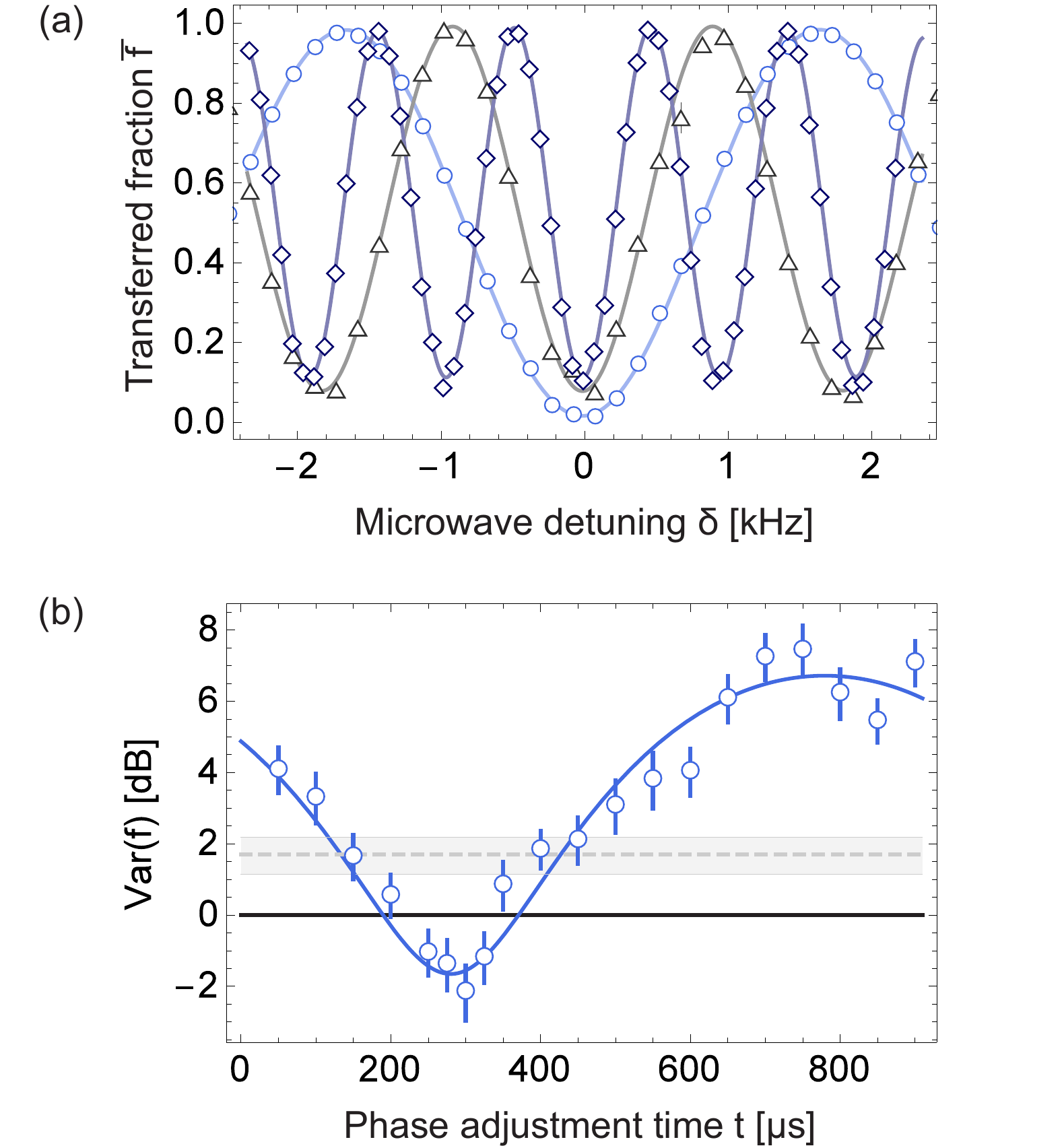}
	\caption{
		\label{fig2}		
		Output of the interferometer.
		(a) The microwave detuning is varied for different Ramsey times $\tau_\mathrm{R}=250\,\mu\mathrm{s}$ (\protect\includegraphics[height=3mm]{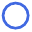}), $\tau_\mathrm{R}=500\,\mu\mathrm{s}$ (\protect\includegraphics[height=3mm]{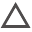}), and $\tau_\mathrm{R}=1000\,\mu\mathrm{s}$ (\protect\includegraphics[height=3mm]{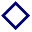}).
		The phase shift, set by the microwave detuning ($x$-axis) and the evolution time $\tau = \tau_\mathrm{R} + \tau_\mathrm{mw}$, results in the Ramsey fringes in the transferred fraction.
		The solid lines (\protect\includegraphics[width=2mm]{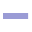}, \protect\includegraphics[width=2mm]{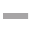}, \protect\includegraphics[width=2mm]{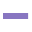}) represent the results of our model.
		(b) The phase adjustment time $t$ is varied and the corresponding variance of the transferred fraction is recorded with respect to shot noise (\protect\includegraphics[height=2ex]{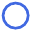}).
		The data is well reproduced by a sinusoidal fit (\protect\includegraphics[width=2mm]{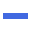}) and reaches clearly below shot noise (\protect\includegraphics[height=2ex]{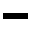}).
		Result of the classical interferometer (\protect\includegraphics[height=2ex]{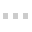}).
		The error bars and shaded areas represent the statistical uncertainty.}  
\end{figure}

Our atomic clock consists of a four-mode linear Ramsey interferometer when described in terms of the Zeeman 
states $\ket{\pm 1} = \ket{F=1,m_F=\pm 1}$, $\ket 0 = \ket{F=1,m_F=0}$  and $\ket{e} = \ket{F=2,m_F=0}$ (see Fig.~\ref{fig1}).
It can however be reduced to a standard two-level Ramsey sequence in terms of the magnetically insensitive clock states $\ket{g} = \tfrac{\ket{+1} + \ket{-1}}{\sqrt{2}}$ and $\ket{e}$ (see Fig.~\ref{fig1} and \cite{Suppl}).
The atoms are prepared in a balanced superposition of these clock states, where they sense the evolution of time by acquiring a phase shift $\theta=2\pi \, \tau \, \delta$, which depends on the detuning $\delta$ of the employed  microwave oscillator and the phase evolution time $\tau = \tau_\mathrm{R} + \tau_\mathrm{mw}$, where $\tau_\mathrm{R}$ is the Ramsey interrogation time and $\tau_\mathrm{mw}$ is the microwave pulse duration.
After a second coupling between the two states, the phase estimation is obtained from the measurement of the fraction $f =\frac {N_{g}} N \approx \frac {N_{+1} + N_{-1}} N$ of atoms in the output state $\ket{g}$, which can be obtained directly from an absorption image of all three Zeeman components. 
Figure \ref{fig2} (a) presents the Ramsey fringes for the classical case, when the hyperfine level $\ket{g}$ is initially empty.
The average fraction $\bar{f}$ is shown as a function of the  microwave detuning for three Ramsey times $\tau_\mathrm{R}$ and is well reproduced by a single-atom model~\cite{Suppl}.
The slightly reduced contrast for larger $\tau_\mathrm{R}$ stems from the influence of a small radio-frequency detuning as well as magnetic field noise.
For small $\tau_\mathrm{R}$ the data are in good agreement with the noiseless prediction $\avef = \sin^2 \pi \delta \tau$. 

The clock performance is evaluated for a vanishing time $\tau_\mathrm{R} = 0$ between the two microwave pulses to minimize technical noise.
For a microwave pulse length of $2 \, \tau_\mathrm{mw} = 90.4\,\mu\mathrm{s}$ and a detuning of $\delta = -5.5\,$kHz we reach the mid-fringe position $\theta = 2\pi \, \tau_\mathrm{mw} \, \delta = \pi/2$, where the slope $\partial \avef /\partial \theta$ reaches its maximum value 1/2.
Here, $\tau_\mathrm{mw}$ is chosen such that all atoms return to the state $\ket{0}$.
On mid-fringe position, the full interferometer sequence can be described as a single, symmetric beam splitter with the output states $\ket{0}$ and $\ket{g}$.
Due to the large number of atoms in the state $\ket{0}$, which act as a local oscillator with a defined phase $\varphi$ in the quantum optics sense, 
the interferometer sequence presents a standard homodyne measurement of the quadratures in state $\ket{g}$.
Therefore, the fluctuations of the interferometer output reflect the quadrature fluctuations: 
$(\Delta f)^2 = (\Delta X)^2 / (2N)$~\cite{Suppl}.
Here, $X = \frac{1}{\sqrt{2}}(e^{-i\varphi} \gm + e^{i\varphi} \gm^\dag)$ and $P = \frac{1}{i \sqrt{2}}(e^{-i\varphi} \gm - e^{i\varphi} \gm^\dag)$ 
are quadrature operators of the symmetric state $\ket{g}$, defined in terms of the creation and annihilation operators $g^\dag$ and $\gm$, respectively, 
and $\varphi$ is the local oscillator phase.
With an initially empty state $\ket{g}$, the quadrature fluctuation is $(\Delta X)^2 = \tfrac{1}{2}$.
This limits the ideal phase estimation uncertainty to $(\Delta \theta)^2 =(\Delta f)^2 / (\partial \avef / \partial \theta)^2 = 1/N$.
In our experiments, we record a value of $(\Delta f)^2 = 1.48 / N $, which is $1.69\,$dB above the vacuum limit
due to technical noise mainly caused by magnetic field fluctuations~\cite{Suppl}.

The sensitivity of our interferometer can surpass the SQL when quadrature fluctuations are squeezed below the vacuum limit, $(\Delta X)^2 < 1/2$.
We create a squeezed vacuum state by initiating spin dynamics in the Bose-Einstein condensate prior to the interferometer sequence.
In direct analogy to optical parametric down-conversion, spin dynamics creates pairs of atoms according to the two-mode squeezing Hamiltonian~\cite{Klempt2010}  $H=\hbar\Omega (a_{+1}^\dag a_{-1}^\dag + a_{+1} a_{-1})$, where $a_{\pm 1}^\dag$ and $a_{\pm 1}$ are the creation and annihilation operators for atoms in $\ket{\pm 1}$, and $\Omega = 2 \pi \times 3.9\,$s$^{-1}$ is the spin dynamics rate. 
Using the operators $g = (a_{+1} +  a_{-1}) / \sqrt{2}$ and $\asm = ( a_{+1} -  a_{-1} )/\sqrt{2}$,
this Hamiltonian simplifies to  $H=H_g - H_{\asm}$, with  $H_{g} = \frac{\hbar\Omega}{2} (\gm^\dag \gm^\dag + \gm \gm )$ and the analogous definition for $H_{\asm}$. 
Spin dynamics $e^{-i H t/\hbar} = e^{- ir (\gm^\dag \gm^\dag + \gm \gm)/2} \otimes e^{ir (\asm^\dag \asm^\dag + \asm \asm)/2}$
can thus be written in terms of the product of usual single-mode quadrature-squeezing operators~\cite{Scully1997}, where $r = \Omega t$.
We apply spin dynamics for $32\,$ms, which creates a mean number $\sinh^2 r=0.75$ of atoms in each of the two states.
This number is extremely small compared to the total of $N\approx 10^4$ atoms, such that the 
influence of the antisymmetric state to the interferometer signal is negligible and we can approximate 
$f= \frac {N_{+1} + N_{-1}} N = \frac {N_g + N_{\asm}} N \approx \frac {N_g} N$.
Even though the symmetric state $\ket {g}$ is only weakly populated, it has a strong influence on the interferometric sensitivity.
The squeezing allows for reduced quadrature fluctuations of $(\Delta X)^2 =\tfrac{1}{2} e^{-2r} < \tfrac{1}{2}$ for $r>0$ 
at an optimal local oscillator phase $\varphi=\tfrac \pi 4$.
Experimentally, the phase $\varphi$ is adjusted by applying a controlled energy shift with the microwave dressing field for a variable duration $t$ prior to the interferometer sequence.

\begin{figure}[ht!]
	\centering
	\includegraphics[width=\columnwidth]{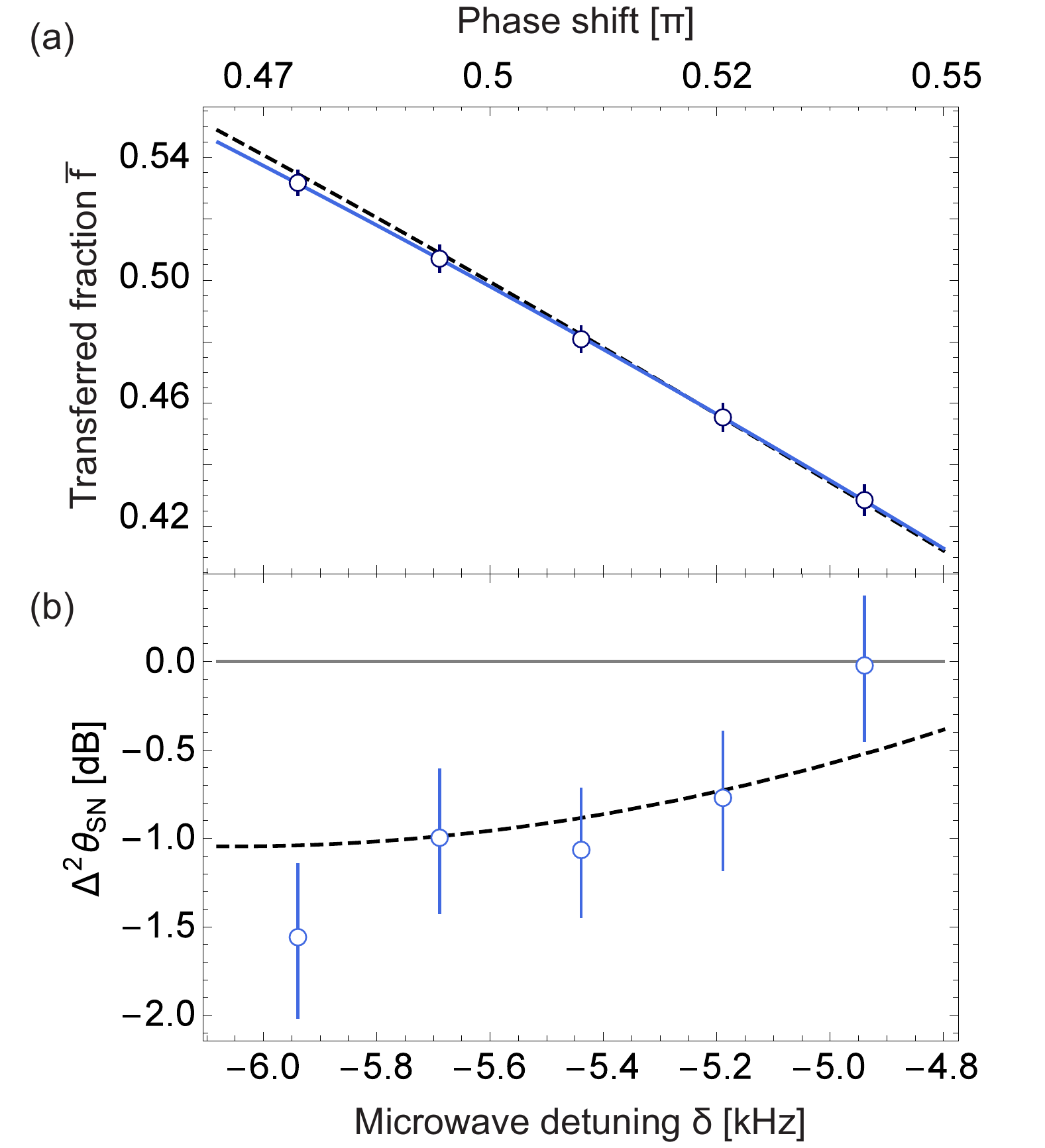}
	\caption{
		\label{fig3}
Determination of the phase estimation uncertainty. (a) Mean values of the transferred fraction $\bar{f}$ (\protect\includegraphics[height=3mm]{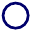}) for five different microwave detunings (bottom $x$-axis) and corresponding phase shifts (top $x$-axis) close to the mid-fringe position. The slope of the linear fit (\protect\includegraphics[height=2ex]{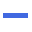}) is close to the optimum, as represented by our model (\protect\includegraphics[height=2ex]{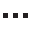}). The error bars represent the (sub-shot-noise) statistical uncertainty. 
(b) Phase estimation uncertainty $\Delta^2 \theta_{\mathrm{SN}} $(\protect\includegraphics[height=3mm]{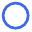}) for the same detuning reach well below shot noise (\protect\includegraphics[width=2mm]{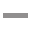}). The small detuning dependency of the recorded sensitivity is reproduced by our single-atom model including magnetic field noise (\protect\includegraphics[height=2ex]{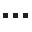}). The best phase estimation uncertainty of $-1.56^{+.41}_{-.45}\,$dB below the SQL is reached at a detuning of $-5.9\,$kHz. Error bars are statistical uncertainty.}
\end{figure}

Figure~\ref{fig2} (b) shows the variance of the population imbalance $(\Delta f)^2$ as a function of the adjusted phase relation. 
At an optimum value of $t = 300\,\mu\mathrm{s}$, a minimal variance of $-2.12 ^{+.70}_{-.83}\,$dB below shot noise is reached. 
Figure~\ref{fig3} (a) shows the fraction $\bar{f}$ as a function of the detuning in the vicinity of the mid-fringe position. 
The slope  is proportional to the contrast of the interferometer and depends on the coherence properties of the input state.
A fit (blue solid line) yields a value of $0.48\,$rad$^{-1}$, which is close to the optimal value of $0.5\,$rad$^{-1}$. 
The variances of the population imbalance and the fitted slope yield the phase estimation uncertainty $(\Delta \theta)^2 =(\Delta f)^2 / (\partial \avef / \partial \theta)^2$ displayed in Fig.~\ref{fig3} (b). 
At a detuning of $-5.9\,$kHz, $(\Delta \theta)^2$ reaches a minimum value $-1.56^{+.41}_{-.45}\,$dB below the SQL.
The two-sample variance, which rejects long term technical drifts and is therefore better suited to estimate the fundamental noise, reaches $-2.05^{+.34}_{-.37}\,$dB below the SQL.

\begin{figure}[ht!]
	\centering
	\includegraphics[width=\columnwidth]{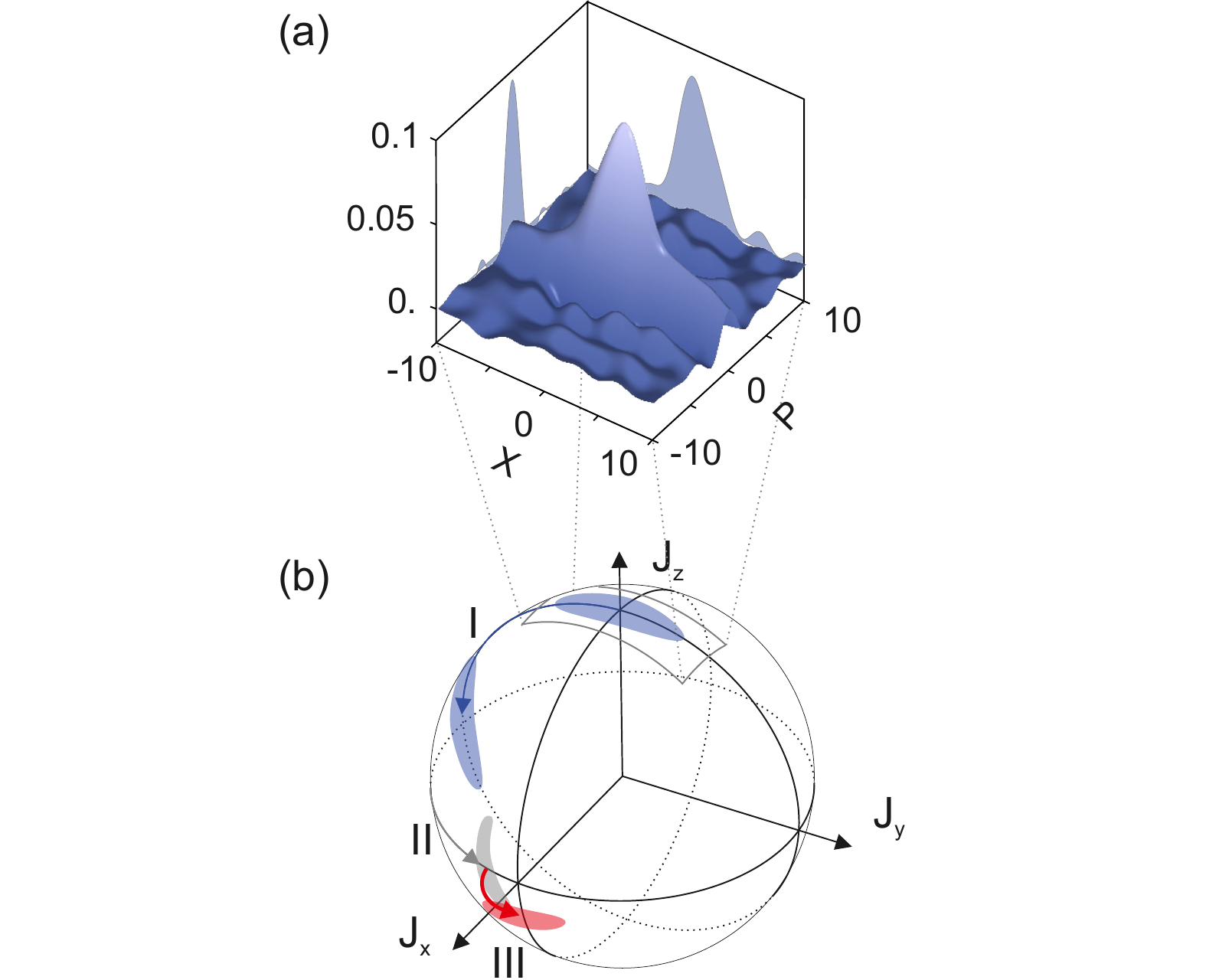}
	\caption{
		\label{fig4}		
		Reconstructed Wigner function and Bloch sphere representation. (a)  The data presented in Fig.~\ref{fig2} (b) is used to reconstruct the two-dimensional Wigner function in the $X$-$P$-space. Here, the Wigner function of the state after the optimal phase adjustment time of $t=300\,\mu\mathrm{s}$ is shown. (b) The interferometer is illustrated on the multi-particle Bloch sphere for the states $\ket{g}$ and $\ket{e}$, in terms of the pseudo-spin operators $J_x=\frac{1}{2}(\eem^\dag \gm + \gm^\dag \eem)$, $J_y=\frac{1}{2i}(\eem^\dag \gm - \gm^\dag \eem)$, $Jz=\frac{1}{2}(\eem^\dag \eem - \gm^\dag \gm)$. The employed squeezed vacuum corresponds to an elliptical uncertainty disk with variable orientation angle depending on the phase adjustment time $t$. An optimal orientation angle, as shown, allows for a measurement of the transferred fraction with a sub-shot-noise uncertainty. The two radio-frequency pulses generate rotations around the $J_x$-axis (I/III). The phase shift corresponds to a rotation around the $J_z$-axis (II). }
\end{figure}

The interferometric measurements also allow for a reconstruction of the squeezed vacuum state in mode $\ket{g}$.
The Wigner function in $X$-$P$-space after the optimal phase adjustment time of $300\,\mu\mathrm{s}$ is obtained from an inverse Radon transformation (Fig.~\ref{fig4} (a) and \cite{Suppl}).
Its profile is very close to the expected Gaussian distribution, and is characterized by the squeezed and anti-squeezed widths along the $X$ and $P$ directions, respectively.
While this single-mode picture successfully describes the physics of our experiments, it can equivalently be described by spin squeezing of the usual two-mode pseudo-spin operators, as visualized on the multi-particle Bloch sphere in Fig.~\ref{fig4} (b). 
It is worth noting, that these collective pseudo-spin operators are identical to the SU(2) subspaces exhibiting spin-nematic squeezing reported in Ref.~\cite{Hamley2012}. 
Furthermore, the created squeezed vacuum state can also be employed for phase sensing in a nonlinear interferometer scheme of the SU(1,1) type~\cite{Linnemann2016,Gabbrielli2015}.

In summary, our experiments present the first proof-of-principle implementation of squeezed vacuum in an atomic microwave clock.
Microwave fountain clocks, providing the realization of the SI second, are currently limited by the SQL~\cite{Santarelli1999,Millo2009,Weyers2009,Dobrev2016}.
In combination with the recently developed sources of Bose-Einstein condensed atoms with small densities~\cite{Muentinga2013, Dickerson2013} and high repetition rates~\cite{Rudolph2015}, our results pave the way for the development of a new generation of atomic microwave clocks operating beyond the SQL~\cite{Suppl}.
Our method is particularly robust during state preparation.
In contrast to existing proposals, it avoids the generation of entangled states with a symmetric population of the two hyperfine levels, which is plagued by two-body losses in the excited hyperfine state.
The limitations of our method for sub-SQL interferometry have not been reached yet: besides overcoming technical restrictions, it has been shown~\cite{Pezze2008} that an optimized version of the present interferometric scheme can reach the ultimate Heisenberg limit of phase sensitivity $\Delta \theta = 1/N$.

\begin{acknowledgments}
We acknowledge support from the Centre for Quantum Engineering and Space-Time Research (QUEST) and from the Deutsche Forschungsgemeinschaft (Research Training Group 1729). We acknowledge support from the European Metrology Research Programme (EMRP) within QESOCAS. The EMRP is jointly funded by the EMRP participating countries within EURAMET and the European Union. L.P. and A.S. acknowledge financial support by the EU-STREP project QIBEC. J.A. acknowledges support by the Lundbeck Foundation.
\end{acknowledgments}

\bibliography{Kruse}

\cleardoublepage

\section{Supplemental Material}

\subsection{Experimental sequence}
\noindent
{\it Squeezed-vacuum state preparation.}
We start the experiments with an almost pure Bose-Einstein condensate of $10,000$ $^{87}$Rb atoms in an optical dipole potential with trap frequencies of 
$2 \pi \times (280,220,180)\,$Hz.
At a homogeneous magnetic field of $2.6\,$G the condensate is 
transferred from the level $\ketF{F,m_F}=\ketF{2,2}$ to the level $\ketF{1,0}$ by a series of three resonant microwave pulses. 
During this preparation, two laser pulses resonant to the $F=2$ manifold clean the system of atoms in unwanted spin states. 
Directly before spin dynamics is initiated, the output states $\ketF{1,\pm 1}$ are emptied with a pair of microwave $\pi$~pulses from $\ketF{1,+1}$ to $\ketF{2,+2}$ and from $\ketF{1,-1}$ to $\ketF{2,-2}$ followed by another light pulse. This cleaning procedure ensures that no thermal or other residual excitations are present in the two states where the squeezed vacuum is prepared, since this may lower the entanglement properties~\cite{Lewis-Swan2013}.

To generate the squeezed vacuum a microwave frequency which is red-detuned by $208\,$kHz to the transition between the levels $\ketF{1,-1}$ and $\ketF{2,-2}$ is adiabatically ramped on within $675\,\mu$s.
The microwave shifts the level $\ketF{1,-1}$ by $500\,$Hz, depending on the chosen detuning, to compensate for the quadratic Zeeman effect of $q=491\,$Hz, such that multiple spin dynamics resonances can be addressed~\cite{Klempt2010,Luecke2014}.
Each resonance condition belongs to a specific spatial mode of the states $\ketF{1,\pm 1}$  to which the atoms are transferred. If the energy of the level $\ketF{1,-1}$ is reduced, more internal energy is released, and higher excited spatial modes are populated~\cite{Klempt2010}.
Here, we choose the first resonance, where spin dynamics leads to a population of the levels $\ketF{1,\pm 1}$ in the ground state of the effective potential.
This ensures an optimal spatial overlap between the atoms in the three contributing levels. This resonance condition is reached, when the input state (two atoms in the BEC in the level $\ketF{1,0}$ at the energy of the chemical potential) is exactly degenerate with the output state (two atoms in the levels $\ketF{1,\pm 1}$ including dressing, trap energy and mean-field shift).
Due to this degeneracy, the phase relation between the initial condensate and the output state stays fixed during the spin dynamics evolution time. For this configuration, we have checked that spin dynamics is the only relevant process which produces atoms in the state $\ketF{1,\pm 1}$ (see Ref.~\citenum{Peise2015}, Fig.~3).
Subsequently, the microwave dressing field is ramped down within $675\,\mu$s, stays off for a variable duration between $25$ and $850\,\mu\mathrm{s}$ before the interferometer sequence is initiated.
The variable hold time $t$ allows for an adjustment of the phase $\varphi$ of the remaining condensate relative to the squeezed vacuum state. 

{\it Interferometer operations.}
We prepare all atoms in the state $\ket 0$ and then apply a resonant radio-frequency pulse with a frequency of $1.835\,$MHz and a duration of $47\,\mu$s, which generates a superposition with $25\,\%$ of the atoms in the levels $\ket{\pm 1}$,  respectively, and $50\,\%$ in the level $\ket{0}$. 
This three-mode operation can be described equivalently as a $\pi/2$ pulse between the initial state $\ket{0}$ and the symmetric state $\ket{g}$. 
The antisymmetric state $\ket{\asm}$ remains unpopulated. 
In a second step, we apply a microwave $\pi$~pulse to the transition $\ket{0} \rightarrow \ket{e}$ with a detuning $\delta$ and a pulse length $\tau_\mathrm{mw}=45.2\,\mu\mathrm{s}$. 
The combination of the radio-frequency pulse and the  microwave $\pi$~pulse realizes an effective $\tfrac \pi 2$~pulse between the symmetric state $\ket{g}$ and the upper magnetically insensitive state $\ket{e}$ --- the first $\tfrac \pi 2$~pulse of the effective two-level Ramsey sequence. 
After a variable Ramsey time $\tau_\mathrm{R}$, a second microwave $\pi$~pulse with the same parameters transfers the atoms back to $\ket{0}$ and a final resonant radio-frequency pulse couples $\ket{g}$ and $\ket{0}$.
Again, the combination of the microwave pulse and the subsequent radio-frequency pulse acts as an effective $\tfrac \pi 2$~pulse between the clock levels $\ket e$ and $\ket g$ and thus concludes the Ramsey sequence.
It converts the acquired phase into a measurable population difference.
This phase $\theta=2\pi \, \tau \, \delta$ is sensitive to both the microwave detuning $\delta$ and the phase evolution time $\tau = \tau_R + \tau_\mathrm{mw}$.
During the complete interferometer sequence, the second-order Zeeman shift is compensated by shifting the state $\ket{1,-1}$ using a microwave coupling to the state $\ket{2,-2}$, which is chosen such that both radio-frequency transitions are resonant but the resonance condition for spin dynamics is not fulfilled. 

{\it Detection.}
After the interferometer sequence, the dipole trap is switched off to allow for ballistic expansion. Following an initial expansion of $1.5\,$ms to reduce the density, a strong magnetic field gradient is applied to spatially separate the atoms in the three Zeeman levels.
Finally, the number of atoms in the three clouds, $N_0$ and $N_{\pm 1}$, is detected by absorption imaging on a CCD camera with a large quantum efficiency. The statistical uncertainty of a number measurement is dominated by the shot noise of the photoelectrons on the camera pixels and amounts to $16$~atoms.
We estimate the uncertainty of the total number calibration to be less than $1\,\%$.

\subsection{Relation between single-mode quadratures and two-mode pseudo-spin operators}
\noindent
Atom interferometers are usually described with pseudo-spin operators, where the atoms in the two interferometer states $\ket{g}$ and $\ket{e}$ are treated as effective spin-1/2 particles.
The action of an interferometer can thereby be described on the multi-particle Bloch sphere, which visualizes the many-particle state in terms of the total spin of the ensemble.
The maximum spin length $J= \sqrt{\tfrac N 2 (\tfrac N 2 +1)}$ is reached for indistinguishable bosons in a single spatial mode.
In this case, the three spin operators can be defined as  $J_x=\frac{1}{2}(\eem^\dag \gm + \gm^\dag \eem)$, $J_y=\frac{1}{2i}(\eem^\dag \gm - \gm^\dag \eem)$, $Jz=\frac{1}{2}(\eem^\dag \eem - \gm^\dag \gm)$, in terms of the creation and annihilation operators of the corresponding states.
How can our experiments with a strongly populated state $\ket{e}$ and a squeezed vacuum in state $\ket{g}$ be described in this picture?
In the limit $N_e \approx N \gg 1$, we can apply the mean-field approximation $\eem, \eem^\dag \approx \sqrt{N}$. 
The spin operators thus become equivalent to the quadratures, $J_x=\sqrt N X$ and $J_y=\sqrt N P$.
For the created state, the collective spin thus points along the $z$-direction, $J_z=N/2$, and is squeezed in the orthogonal plane in the same way as the quadratures.
Figure~4 (b) indicates the fluctuations of the orthogonal spin components $J_x$ and $J_y$, which are well represented by the reconstructed Wigner function in the mean-field approximation, visualizing the close connection between spin-squeezing and quadrature squeezing~\cite{DuanPRA2002, Wang2003}.
The figure also presents the action of the interferometer sequence which results in a phase estimation uncertainty beyond the SQL.
It is worth noting, that these collective pseudo-spin operators are identical to the $\{S_x,Q_{yz},(Q_{zz}-Q_{yy})\}$ subspace exhibiting spin-nematic squeezing reported in Ref.~\citenum{Hamley2012}.
The $\{S_y,Q_{xz},(Q_{xx}-Q_{zz})\}$ subspace corresponds to the pseudo-spin basis of the states $\ket e$ and $\ket{\asm}$ and exhibits squeezing in the orthogonal direction.

\subsection{Theoretical description of the three-mode interferometer}
\noindent
The symmetric rf coupling between $\ket{0}$ and $\ket{\pm1}$ is described by the Hamiltonian~\cite{Peise2015a, Gross2011} 
$\Hrf = \frac{\hbar \Omega_{\rm rf}}{2\sqrt{2}} \big( \ops^\dag \opp + \ops \opp^\dag + \opi^\dag \opp + \opi \opp^\dag \big)$.
This operation, implemented for a time $t = \pi/2 \times \Omega_{\rm rf}^{-1} $ and followed by a microwave $\pi$ pulse on the clock transition coupling $\ket{0}$ to $\ket{e}$, 
corresponds to a balanced beam splitter $\Ubs = e^{-i (\pi/2) J_x}$ between $\ket{e}$ and the symmetric state $\ket{\gm}$.
Notice that the antisymmetric state $\ket{\gbm}$ is untouched by this operation.
Free precession leads to the accumulation of a relative phase $\theta$ between the modes $\ket{e}$ and $\ket{\gm}$.
This is described by $\Ups = e^{-i \theta \hat{J}_z}$. 
The interferometer is closed by a second beam splitter (i.e. a microwave $\pi$ pulse followed by a balanced rf coupling) 
such that the overall transformation is $\hat{U}_{\rm clock} = \Ubs \Ups  \Ubs = e^{-i \pi J_x }e^{-i \theta J_y}$,
corresponding (up to $e^{-i \pi J_x }$) 
to a spin rotation of an angle $\theta$ around the $y$-axis. 
The interferometer signal is the average fraction of atoms in the $m_F = \pm 1$ levels:
\begin{equation} \label{interf}
f = \frac {N_{+1} + N_{-1}} N = \frac {N_g + N_{\asm}} N.
\end{equation}
Taking into account the above interferometer transformation, we have 
\begin{equation} \label{feq}
f = \frac{1}{2} - \frac{J_z}{N}  \cos \theta - \frac{J_x}{N}  \sin \theta +  \frac{N_{\asm}}{2N}.
\end{equation}
In the limit $N \gg N_g, N_{\asm}$, neglecting fluctuations of the total number of particles, we find 
$\avef = \sin^2 \tfrac{\theta}{2}$ (taking $X=0$) and, 
neglecting fluctuations of the total number of particles, 
\be
(\Delta f)^2 \approx \frac{(\Delta X)^2}{2N} \sin^2 \theta.
\ee
The phase sensitivity reaches its maximum at mid-fringe ($\theta=\pi/2$) where
\be
(\Delta \theta)^2 = \frac{(\Delta f)^2}{ (\partial \bar{f} / \partial \theta)^2} \bigg\vert_{\theta=\pi/2} = \frac{2}{N}(\Delta X)^2.
\ee

\subsection{Single-atom model}

The Hamiltonian of a single particle under the influence of a radio-frequency coupling of the levels $\ketF{F,m_F}=\ketF{1,0} \leftrightarrow \ketF{1,\pm 1}$ and a microwave coupling of the levels $\ketF{1,0} \leftrightarrow \ketF{2,0}$ can be written in matrix form as
\begin{equation}
H = \left(
\begin{array}{c c c c}
\delta_\mathrm{rf} + \tfrac q 2 & \tfrac{\Omega_\mathrm{rf}}{2\sqrt{2}} & 0 & 0\\
\tfrac{\Omega_\mathrm{rf}^*}{2\sqrt{2}} & 0 & \tfrac{\Omega_\mathrm{rf}^*}{2\sqrt{2}} & \tfrac{\Omega_\mathrm{mw}}{2}\\
0 & \tfrac{\Omega_\mathrm{rf}}{2\sqrt{2}} & -\delta_\mathrm{rf} + \tfrac q 2 & 0\\
0 & \tfrac{\Omega_\mathrm{mw}^*}{2} & 0 & -\delta
\end{array}
\right)
\end{equation}
in the basis corresponding to the levels $[ \ketF{1, +1},  \ketF{1,0},  \ketF{1, -1},  \ketF{2, 0} ]$.
Here the parameters are the coupling strength $\Omega_\mathrm{rf}$ and the detuning $\delta_\mathrm{rf}$ of the applied radio-frequency field and the coupling strength $\Omega_\mathrm{mw}$ and the detuning $\delta$ of the applied microwave field as well as the energy difference $q$ of the symmetric state with respect to the level $\ketF{1,0}$.
It is straight forward to obtain the unitary evolution of a single-particle wave function as $U(t) = \exp(-i t H)$.
For instance, to obtain the unitary evolution during a radio-frequency pulse of length $\tau_\mathrm{rf}$, we set the microwave coupling to zero $\Omega_\mathrm{mw}=0$ and compute $U_\mathrm{rf}(\tau_\mathrm{rf}) = \exp(-i \tau_\mathrm{rf} H)$.
In this way, we can simulate the classical interferometer.

\begin{figure}[ht!]
	\centering
	\includegraphics[width=\columnwidth]{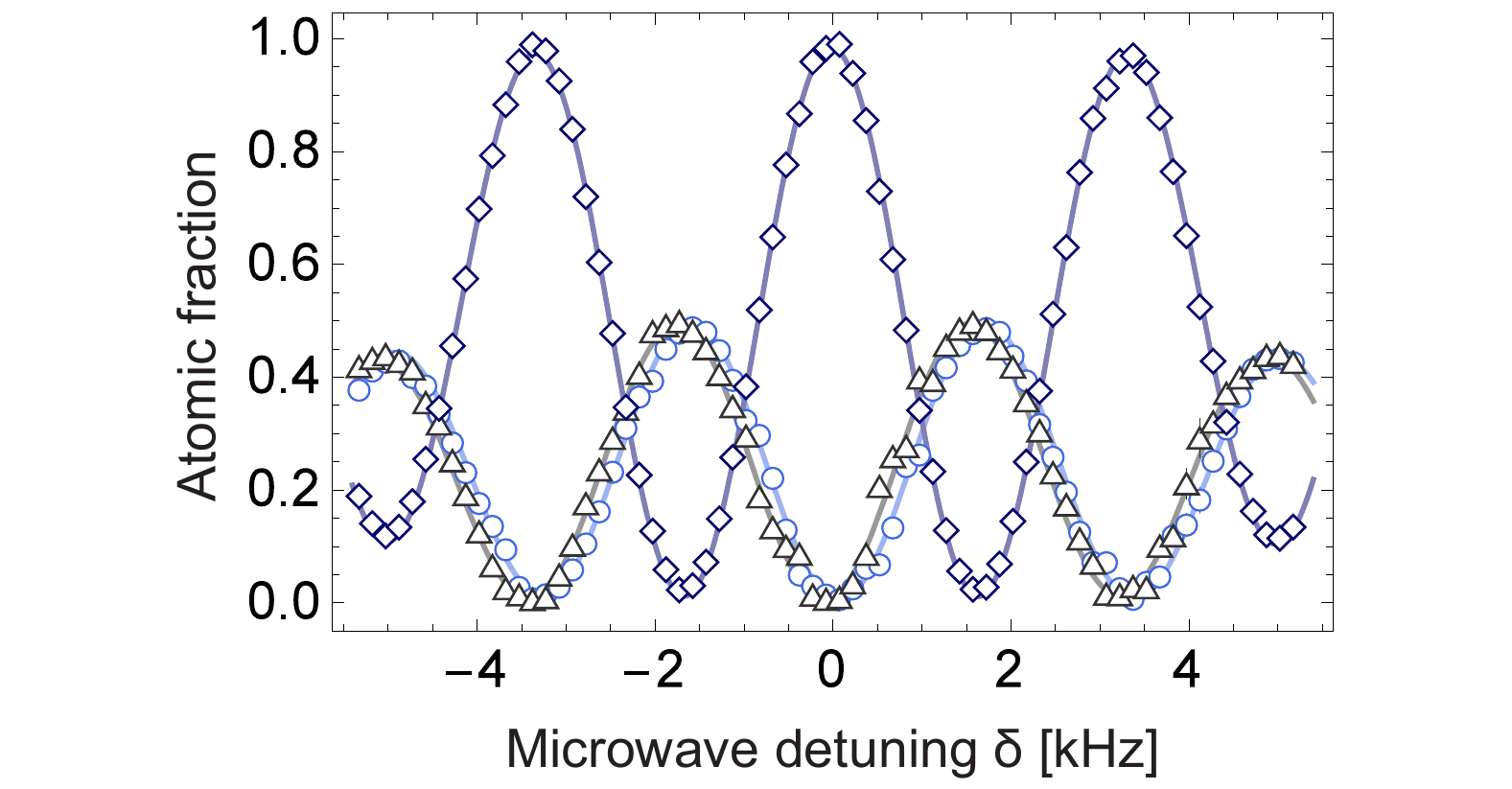}
	\caption{
		\label{figSup2}		
		Atomic fraction in the Zeeman levels. The microwave detuning is varied for a Ramsey time $\tau_\mathrm{R}=250\,\mu\mathrm{s}$. The atomic fractions in the levels $\ketF{1,-1}$ (\protect\includegraphics[height=3mm]{icon-2a-1}), $\ketF{1,1}$ (\protect\includegraphics[height=3mm]{icon-2a-2}) and $\ketF{1,0}$ (\protect\includegraphics[height=3mm]{icon-2a-3}) are recorded. The solid lines (\protect\includegraphics[width=2mm]{line-2a-1}, \protect\includegraphics[width=2mm]{line-2a-2}, \protect\includegraphics[width=2mm]{line-2a-3}) represent the results of our model.}  
\end{figure}

In a Ramsey sequence, a fraction $P_{\pm 1}$ of the remaining atoms in the level $\ketF{2,0}$ is transferred to the levels $\ketF{2,\pm 1}$ by the second radio-frequency coupling.
These atoms are indistinguishable from the atoms in $\ketF{1,\mp 1}$ for our absorption detection since they experience almost the same acceleration due to the magnetic field gradient.
In order to get good agreement with the experimental data, we thus add $P_{\pm 1} N_e$ to the number of atoms $N_{\mp 1}$ obtained from our single-particle simulation and $(1-P_{+1}-P_{-1}) N_e$ to $N_0$.
The model is extended to a Monte Carlo simulation to include the magnetic field noise by repeating the calculation for a random set of magnetic fields.
The parameters have to be adapted accordingly for each calculation.
In the same way, other noise sources like radio-frequency or microwave power fluctuations are simulated in a detailed noise analysis.
The coupling strengths, the magnetic field noise and the energy difference $q$ are measured independently and we achieve excellent agreement with the experimental data with the radio-frequency detuning as the only free parameter as presented in Fig.~2 (a) and Fig.~\ref{figSup2}.
The radio-frequency detuning leads to the small phase shift between the oscillation of the fraction of atoms in the $m_\mathrm{F} = +1$ state and the fraction of atoms in the $m_\mathrm{F} = -1$ state as can be seen in Fig.~\ref{figSup2}.
This phase shift is due to a mean population of the antisymmetric state.
Hence, a radio-frequency detuning can lead to a reduced contrast of our interferometer.
However, it is important to notice that it does not shift the fringe positions and does thus not effect the accuracy of the clock.

\subsection{Noise analysis}
\noindent
The additional technical noise of the interferometer is due to the combination of several noise sources.
Most importantly, shot-to-shot fluctuations of the magnetic field lead to a differential phase shift $\delta \phi = t \, \Delta B \, 700\,\mathrm{kHz/G}^2$ of the $m_F = \pm 1$ levels.
In the basis of the symmetric and antisymmetric states, this drives the initially purely symmetric state into a superposition of the symmetric and the antisymmetric state.
\begin{eqnarray}
\ket g &=& \frac 1 {\sqrt 2} (\ket{1,+1} + \ket{1,-1}) \\ \nonumber
&\to & \frac 1 {\sqrt 2} (e^{i\delta \phi} \ket{1,+1} + e^{-i\delta \phi}\ket{1,-1}) \\ \nonumber
&=& \cos(\delta \phi) \ket g + i \sin(\delta \phi) \ket{\asm}
\end{eqnarray}
The population of the antisymmetric state thus varies as a consequence of fluctuations of the magnetic field, adding noise to the interferometer signal in equation~(\ref{interf}).
This effect is included in the single-particle model described in the previous section, which reproduces the decreasing trend of the fluctuations towards larger detuning for a magnetic field noise of $100\,\mu$G as shown in Fig.~3 (b).
This can be understood in the following way:
The interferometer signal can be expressed as
\begin{eqnarray}
f &=& \frac {N_g + N_{\asm}} N
= \frac { f_\mathrm{id} (N - N_{\asm}) + N_{\asm}} N\\ \nonumber
&=& f_\mathrm{id} + (1-f_\mathrm{id}) \frac{N_{\asm}} N
\end{eqnarray}
where $f_\mathrm{id} = N_g / N$ is the ideal interferometer signal for a vanishing population of the antisymmetric state.
Thus, the fluctuations of the interferometer signal $\Delta f$ due to a fluctuating population of the antisymmetric state $\Delta N_{\asm}$ are
\begin{equation}
\Delta f = (1-f_\mathrm{id}) \frac {\Delta N_{\asm}} N
\end{equation}
Since $1 - f_\mathrm{id}$ decreases for an increasing detuning, this noise contribution becomes increasingly small.

\begin{figure}[ht!]
	\centering
	\includegraphics[width=\columnwidth]{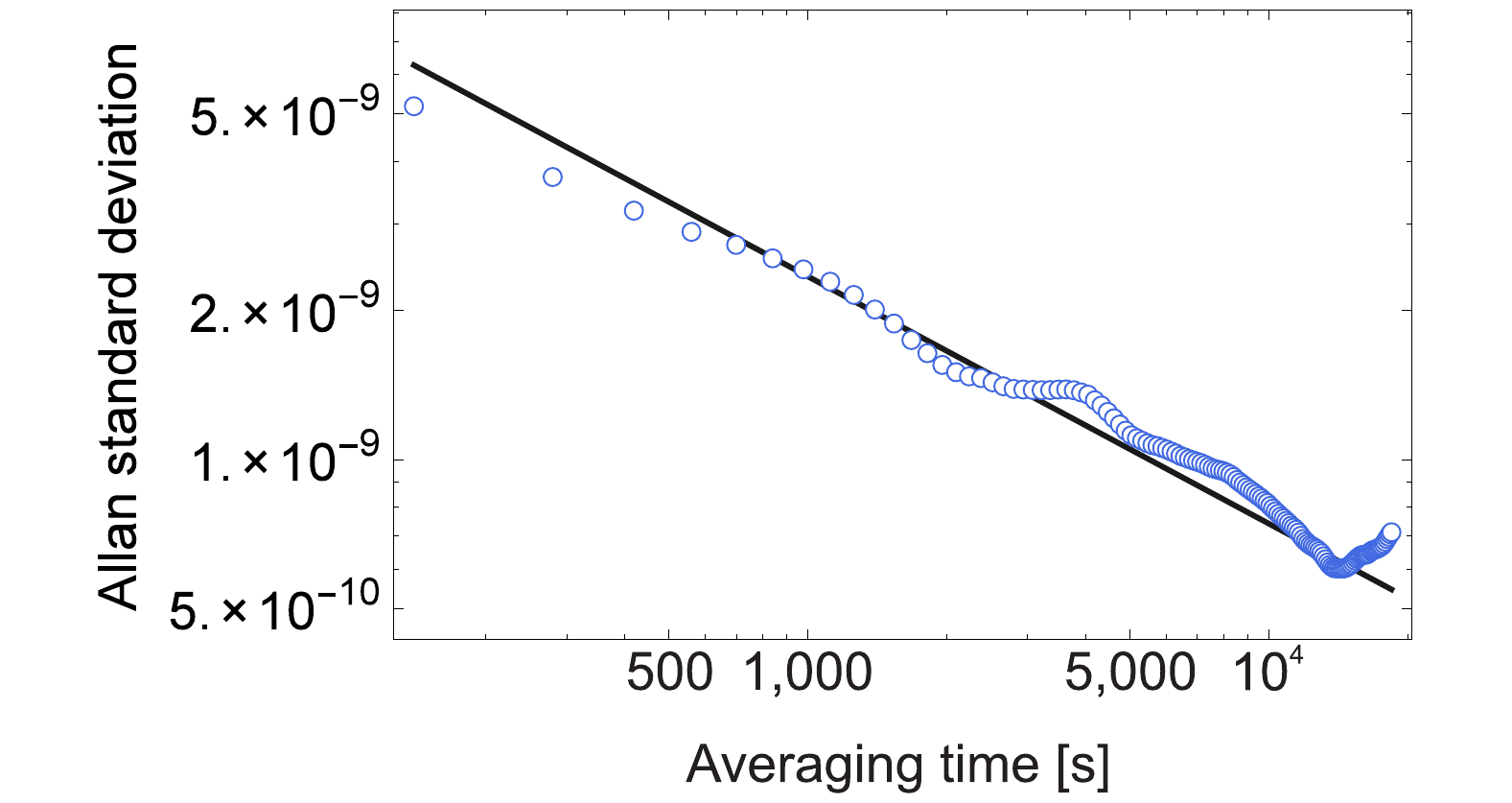}
	\caption{
		\label{figSup1}		
		Allan standard deviation of the fractional instability. The data (\protect\includegraphics[height=3mm]{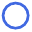}) represent the fractional uncertainty of the frequency measurement as displayed in Fig.~3, as a function of the total averaging time. The instability starts well below the standard quantum limit (\protect\includegraphics[width=2mm]{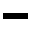}) until it senses environmental noise (see text). }
\end{figure}

The long term noise properties of our measurements are studied in Fig.~\ref{figSup1} which shows the Allan standard deviation of the difference between the atomic frequency and the microwave reference frequency, corresponding to the measurement point at $\delta = -5.9\,$kHz in Fig.~3.
For short times the instability stays below the standard quantum limit, averaging according to $\tau^{-1/2}$.
For longer averaging times $\tau$ the influence of the laboratory's air conditioning cycle at $1000\,$s and $4000\,$s becomes apparent.
At the longest averaging times, the measurement starts sensing fluctuations with the daytime which have a typical time scale of $2 \times 10^4\,$s in our lab.
At the minimum, we reach a relative instability of $6.01 \times 10^{-10}$.
The long-term technical noise sources limit the achievable squeezing in our setup, but could be largely suppressed in a designed metrology apparatus such as a microwave fountain clock. 
However, the two-point variance (the first data point) shows that the squeezing remains stable for large averaging times of more than $10^4\,$s.

\subsection{Wigner reconstruction}
\noindent
The data of Fig.~2 (b) can be interpreted as the marginal distributions of the squeezed vacuum state after homodyne detection.
To show this we introduce new mode operators $\tilde{g} = e^{-i\varphi} g$, where we have absorbed the local oscillator phase.  
The Hamiltonian $H_g$ and the single-mode quadrature-squeezing operators $S_g(i\xi)$
are thus $H_{\tilde{g}} = \tfrac{\hbar \Omega}{2}(\tilde{g}^\dag \tilde{g}^\dag+\tilde{g}\tilde{g})$ and 
$S_{\tilde{g}}(ir)$, respectively, which are formally independent of the local oscillator phase.
In the new state $\tilde{g}$, we can write $X = \tilde{X} \cos \varphi + \tilde{P} \sin \varphi$. 
According to equation~(\ref{feq}), at mid-fringe position and in the limit $N \gg N_\asm$, 
measurements of $f$ corresponds to tomographic quadrature measurements taken at different angles $\varphi$.
By repeating  the quadrature measurements $\sim 100$ times for different values of $\varphi$ (we considered $20$ local oscillator phases in a $\pi$ interval) we recover the probabilities $P(X \vert \varphi)$.
The data is used to reconstruct the Wigner distribution of the squeezed vacuum state via an inverse Radon transformation~\cite{Leonhardt1997, Lvovsky2009}
\be
W( \tilde{X},  \tilde{P}) =  \int d X \int \frac{d \varphi}{2\pi^2} \, P(X\vert \varphi) K( \tilde{X} \cos \varphi + \tilde{P} \sin \varphi - X)
\ee
with the integration kernel 
\be
K(x) = \frac{1}{x^2} [\cos(k_c x) + (k_c x) \sin (k_c x)-1]
\ee
and $k_c$ is a cutoff that reduces numerical artifacts associated with the reconstruction.
Figure 4 (a) shows the reconstructed Wigner function obtained with $k_c=2$.
We used only bare data without any assumption on the quadrature distributions. 
Negative parts are due to the low statistics. 
We have also performed a density matrix reconstruction using a maximum likelihood method~\cite{Peise2015a, Lvovsky2009}
and assuming Gaussian distributions for the quadrature at each $\varphi$.
In this case we obtain a fully positive Wigner function with the same Gaussian shape and quadrature squeezing
of the ones in Fig.~4 (a). 

\subsection{Application to atomic fountain clocks}
\noindent
Today's best microwave clocks, providing the realization of the SI second, are based on laser cooled atoms which are interrogated during free fall in an atomic fountain.
Their statistical uncertainty is currently limited by the SQL~\cite{Millo2009,Weyers2009, Dobrev2016}.
A reduction of the SQL contribution by a mere increase of the number of atoms is problematic:
The increased density leads to a larger collisional shift which gives rise to one of the largest systematic uncertainties.
Our experimental results point the way towards a new generation of microwave clocks.
With the recently demonstrated production of diluted BECs with $1\,$nK kinetic temperature~\cite{Muentinga2013, Dickerson2013}
and $4 \times 10^5$ atoms within $1.6\,$s~\cite{Rudolph2015}
, sources of condensed atoms have been developed that fulfil the requirements of atomic fountain clocks.
Bose-Einstein condensates offer an unprecedented control over the spatial mode of the atoms.
Can atomic microwave clocks benefit from these advances like they did in the past from laser cooling to microkelvin temperatures?
In principle, the spatial control allows to decrease systematic uncertainties due to the quadratic Zeeman effect and due to a disturbed microwave cavity phase with higher precision.
The operation with squeezed vacuum, as demonstrated in our experiments, allows to push  the stability beyond the SQL, thus allowing for short averaging times at small atomic densities and for an improvement of the resulting accuracy. 

\end{document}